\documentclass[12pt]{article}

\usepackage{amsmath, amssymb}
\usepackage{pstricks}
\usepackage{subfig}
\usepackage{graphics}
\usepackage{graphicx}

\usepackage{pstricks}                       
\usepackage{pstricks-add}               
\usepackage{pst-all,pst-math,pst-plot}  
\usepackage{textcomp}
\usepackage[tight,nice]{units}

\def\length{\hbox{\sf length}}

\def\giant{\hbox{\sf giant}}
\def\Exp{{\rm I\!E}}
\def\THEN{\Longrightarrow}
\def\+{ {\oplus} }
\def\-{ {\ominus} }
\def\J{{\cal J}}
\def\M{\cal M}
\def\A{\cal A}
\def\P{\mathbb P}

\def\HB#1{\hbox{~~~#1~~~}}

\def\Z{{\mathchoice
{\hbox{$\sf\textstyle         Z\kern-0.4em Z$}}
{\hbox{$\sf\textstyle         Z\kern-0.4em Z$}}
{\hbox{$\sf\scriptstyle       Z\kern-0.3em Z$}}
{\hbox{$\sf\scriptscriptstyle Z\kern-0.2em Z$}}
}}

\def\HEAD{\small\kern100pt%
\underline{\sl\quad File {\sf WalkAC6.tex}
on {\sf\today~} on {\bf\pageref{LastPage}} pages \quad}}

\newtheorem{theorem}{Theorem}
\newtheorem{lemma}{Lemma}

\begin{document}

\begin{center}
\vspace*{0pt}
{\Large\bf Mean  time  of  archipelagos in 
$1D$ probabilistic cellular automata has phases }\\[10pt]
{\large\bf A. D. Ramos\footnote{Federal University 
of Pernambuco, Department of Statistics, Recife, PE, 50740-540, Brazil;
E-mail: {\bf alex@de.ufpe.br}}}
\end{center} 

\centerline{\bf Abstract}

We study a non-ergodic one-dimensional probabilistic cellular automata, 
where each component can assume the  states $\+$ and $\-.$ 
We obtained  the limit distribution for a set of measures on $\{\+,\-\}^\Z.$ 
 Also, we show that for certain parameters 
of our process  the mean time of convergence can be finite
or infinity. When it is  finite
we have showed that the upper bound is function of the initial distribution.  
\vspace*{5pt}

{\bf Keywords:} {\it particle process,  phase transition, Birth and Death process.}
\section{Introduction}

Generally the theoretical studies about probabilistic cellular automata
or just PCA by simplicity, 
focuses attention to obtain condition under which the PCA is 
non-ergodic or ergodic\cite{Stable} i.e. the process can
keep some knowledge about their initial condition forever;
as opposed to ergodic ones which forget everything
about their initial condition as $t\to\infty$.
 At another direction, when the PCA exhibits non-ergodicity 
we try to characterize the non-trivial invariant measure\cite{DM}.

How long time one  random processes can remember
something about their initial conditions is a important characteristic. 
Let us denote this time by $\tau_{\mu}$ where $\mu$ is the initial
distribution of our process(below we shall define this time at a more formal way).
When the PCA is non-ergodic we have  computational  and theoretical 
 works\cite{Sa, Relaxa} which  describe the expectation of $\tau_{\mu}$
at finite space. 

Even at non-ergodic PCA, understand the behavior of the process for certain initial
conditions is fruitful \cite{Claussen,Macfarlane}. In this work,
for a  set of initial distributions, whose elements we call archipelagos, 
we have shown that  our process converge, we exhibits the 
 limit distribution and  the expectation of $\tau_{\mu}$.
Considering a subset of  archipelagos, which we call archipelago 
of pluses(respectively archipelago of minuses) we get that the expectation of $\tau_{\mu}$
can be finite or infinity. On the first case, expectation of $\tau_{\mu}$ finite, 
we describe the upper bound  this quantity. Also, we get proved that the upper bound
 is  function of the initial distribution. 

  \section{Definitions and Theorems}

We study a  random operators
with one and the same configuration space $\Omega=\{\-,\+\}^{\Z}$ where $\Z$ 
is the set of integer numbers and  
$\-$ and $\+$ are called {\it minus} and {\it plus} respectively. 
A {\it configuration} is an bi-infinite  sequence 
of minuses or pluses. The {\it configuration space} $\Omega$ is 
the set of configurations. Any configuration $x \in \Omega$ 
is determined by its components $x_i$ for all $i \in \Z$.
The configuration, all of whose components are minuses, is called
``{\it all minuses}''. Also, The configuration, all of whose 
components are pluses, is called ``{\it all pluses}''. 

Two configurations $x$ and $y$ are called \textit{close} 
to each other if the set $\{i \in \Z : x_i \neq y_i\}$ is  finite. 
A configuration is called an \textit{island of pluses} if it is close to ``all minuses'',
we denote the set of island of pluses $\Delta_{\+}$.
Respectively a configuration is called an \textit{island of minuses} if it is close to ``all pluses'',
we denote the set of island of minuses $\Delta_{\-}$.
If $x\in\Delta_{\+}$ there are positions
$i < j$ such that $x_{i+1} =x_{j-1}= \+$  and $x_k =\- $ if $k\le i$ or $j\le k$ and for those
same positions $i$ and $j$ we say that a island has {\it length} $ j -i-1$, we denote that quantity
\length($x$).
If $y\in\Delta_{\-}$ there are positions
$i < j$ such that $y_{i+1} =y_{j-1}= \-$  and $y_k =\+ $ if $k\le i$ or $j\le k$ and for those
same positions $i$ and $j$ we say that a island has length $ j - i-1$, we denote that quantity
\length($y$). We denote $\Delta=\Delta_{\-}\cup\Delta_{\+}$ the {\it space of islands}.

The normalized measures concentrated 
in the configuration ``all minuses" and ``all pluses" are denoted by $\delta_{\-}$
and $\delta_{\+}$ respectively. Also, given configuration $x$
we denote the normalized measure concentrated in $x$ by $\delta_x.$

We define {\it cylinders} in $\Omega$ in the usual way. 
By a {\it thin cylinder} we denote any set 
\[
\{x \in \Omega : x_{i_1} = a_1, \dots x_{i_k} = a_k\},
\]
where $a_1,\dots a_k \in \{\-,\+\}$ are parameters.
Thus defined thin cylinder is called a {\it segment cylinder} 
if the indices $i_1,\dots,i_k$ form a segment in $\Z$.
We denote by $\M$ the set of normalized measures on 
the $\sigma$-algebra generated by cylinders in $\Omega$. 
By convergence in $\M$ we mean convergence on all thin cylinders.

We denote by $\A,~\A_{\+}$ and
$\A_{\-}$ the set of normalized measures on 
the $\sigma$-algebra generated by cylinders in $\Delta,~\Delta_{\+}$ 
and $\Delta_{\-}$ respectively.
Any $\mu\in\A$ we call {\it archipelago }.
Any $\mu\in\A_{\+}$ we call  a {\it archipelago of pluses} 
and any $\mu\in\A_{\-}$ we call {\it archipelago of minuses}.

Any map $P : \M \to \M$ is called an {\it operator.} 
Given an operator $P$ and an initial measure $\mu \in \M,$ the resulting 
{\it process} is the sequence $\mu,\ \mu P,\ \mu P^2,\ \ldots.$ 
We say that a measure $\mu$ is {\it invariant} to $P$ if
$\mu P=\mu.$

An arbitrary cellular automaton $P$ is determined by transition 
probabilities $\theta (b_k | a_{k-p},\dots, a_{k+q}) \in [0,1]$, 
where $p,~q$ are non-negative integer numbers, provided for each $k$, 
\[
       \forall\ a_{k-p}, \dots, a_{k+q} \in \{\-,\+\} : \sum_{b_k \in \{\-,\+\}} 
		\theta (b_k | a_{k-p}, \dots, a_{k+q}) = 1.
\]

The following equations give values of $\mu P$ for any $\mu \in \M$ on all segment 
cylinders as linear combinations of the values of $\mu$ on some segment cylinders:
\begin{eqnarray}
     && \forall\ [i,j] \subset \Z,\quad \forall\ a_i, \dots , a_j \in \{\-,\+\} : 
	      \mu P \left( y_k = b_k,~ k \in [i,j] \right) =                  \nonumber\\[10pt] 
     && \hskip -40pt \sum_{a_{i-p},\dots,a_{j+q} \in \{\-,\+\}}
		    \hskip -30pt \mu \left( x_k = a_k,~k \in [i-p,j+q] \right)
        \prod_{k \in [i,j]} \ \theta (b_k | a_{k-p}, \cdots, a_{k + q}). \label{CA}
\end{eqnarray}
Thus a general operator $P$ is defined by (\ref{CA}). 

Now, let us consider a  probabilistic cellular automata in $\Z$,
 which we denote by $F$.
Our operator is defined as follows:
let $p =0 $ and $ q = 1,$ and transition probabilities
\begin{equation}
\begin{array}{cc}
\theta(\+|\-\-)=0;&\theta(\+|\+\-)=\beta;\\
\theta(\+|\-\+)=\alpha;&\theta(\+|\+\+)=1.
\end{array}
\label{eq:transi}
\end{equation}
And $\theta(\-|a_0a_1)= 1 - \theta(\+|a_0a_1)$.
Thus, we have defined our operator.

Evidently $\delta_{\-}$ and $\delta_{\+}$ are invariant measures of our process.
Hence, for $\lambda\in[0,1]$, $\pi_{\lambda}=(1-\lambda)\delta_{\-}+\lambda\delta_{\+}$ 
is invariant to our process. 

Given    $\mu\in\A$, we define
 the random variable
\[
\tau_{\mu}=\inf\{t\ge0:~\mu F^t=\pi_{\lambda} \HB{ for }\lambda\in[0,1]\}.
\]
The infimum of the empty set is $\infty.$

If $\mu\in\A$  we call {\it giant}
of $\mu$ and we denote by \giant($\mu$) the  greatest length of those islands
whose  the $\delta-$measures of the convex combination of $\mu$ are concentrated.
(For the \giant 's definition we are using the result stated in lemma \ref{lem:countable} 
that $\Delta$ is countable).   
If there is not such greatest length, we say that \giant($\mu$)$=\infty.$

We say that our operator $F$ is {\it eroder of archipelago of pluses  in mean linear
time}(respectively {\it eroder of archipelago of minuses  in mean linear
time})
if fixed $\alpha$ and $\beta$ there is constant
$k$ such that
\[
\Exp(\tau_{\mu})\le k(1+\giant(\mu)) ,
\]
 for all  $\mu\in\A_{\+}$ (respectively for $\mu\in\A_{\-}$) whose $\giant(\mu)$ is finite.

Now, we shall declare our main results.

\begin{theorem}
Lets $\alpha>0,~\beta<1$ be. If $\mu\in\A$, then exist $\lambda\in[0,1]$ such that
\[
\lim_{t\to\infty}\mu F^t=\pi_{\lambda}.
\] 
In particular, if  $\mu\in\A_{\+}$ (respectively $\mu\in\A_{\-}$) then
$\lambda=0 $ i.e  $\mu F^t$ goes to $\delta_{\-}$ 
( $\lambda=1 $ i.e  $\mu F^t$ goes to $\delta_{\+}$)
when $t\to\infty$
\label{TEO10}
\end{theorem}

\begin{theorem}Lets $\beta<1,~\mu\in\A_{\+}$ and $\giant(\mu)$ finite. 

(A.\ref{TEO20}) If $\alpha< 1-\beta$ then $\Exp(\tau_{\mu})<\infty;$

(B.\ref{TEO20}) If $\alpha\ge 1-\beta$ then $\Exp(\tau_{\mu})=\infty.$

\label{TEO20}
\end{theorem}

\begin{theorem}Lets $\alpha>0,~\mu\in\A_{\-}$ and $\giant(\mu)$ finite. 

(A.\ref{TEO30}) If $\alpha> 1-\beta$ then $\Exp(\tau_{\mu})<\infty;$

(B.\ref{TEO30}) If $\alpha\le 1-\beta$ then $\Exp(\tau_{\mu})=\infty.$

\label{TEO30}
\end{theorem}

\begin{theorem}Lets $\alpha>0$ and $\beta<1$ be

(A.\ref{TEO40})If $\alpha<1-\beta$ then $F$ is eroder of archipelago of pluses in mean linear
time; 

(B.\ref{TEO40})If $\alpha>1-\beta$ then $F$ is eroder of archipelago of minuses in mean linear
time.
\label{TEO40}
\end{theorem}  

\section{Order}

As very intuitive we shall assume $\-\prec\+$.  
Now, let us introduce a partial order on $\{\+,\-\}^{\Z}$ by saying
that configuration $x$ {\it preceeds} configuration $y$
or, what is the same, $y$ {\it succeeds} $x$ 
and writing $x \prec y$ or $y \succ x$ if
$x_i \le y_i$ for all $i \in \Z$.

Let us say that a measurable set $S \subset \{\+,\-\}^{\Z}$ 
is {\it upper} if
\[
  (x \in S \HB{ and } x \prec y) \ \THEN \ y \in S.
\]
Analogously, a set $S$ is {\it lower} if
\[
  (y \in S \HB{ and } x \prec y) \ \THEN \ x \in S.
\]
It is easy to check that a complement to an upper set is
lower and vice versa.

We introduce a partial order on $\M$ by saying that a normalized
measure $\mu$ {\it preceeds} $\nu$ (or $\nu$ {\it succeeeds} $\mu$) 
if $\mu(S) \le \nu(S)$ for any upper $S$ 
(or $\mu(S) \ge \nu(S)$ for any lower $S$, which is equivalent).

We call an operator $P : \M \to \M$ {\it monotonic} 
if $\mu \prec \nu$ implies $\mu P \prec \nu P$.

The lemma \ref{lem:monoP}  was described in \cite{SCS,ATC} pages 28 and 81 respectively.

\begin{lemma}
Lets $x,~ y$ two configuration.  An operator  $P$  on $\{\+,\-\}^{\Z}$ with transition of probabilities
$\theta_k(.|.)$ is monotonic if only if 
\begin{equation}
x\prec y\THEN\theta_k(\+|x_{k-p}\ldots x_{k+q})\le \theta_k(\+|y_{k-p}\ldots y_{k+q}).
\label{eq:mono}
\end{equation}
\label{lem:monoP}
\end{lemma}

\begin{lemma}
Our operator $F$ is monotonic.
\label{lemm:monoF}
\end{lemma}
{\bf Proof.} It is enough use the lemma \ref{lem:monoP} and the definition (\ref{eq:transi}).

\section{Proof of theorem \ref{TEO10}}

We say that a configuration $x$ is a $(\+\-,i)-${\it jump} if there is position $i$ 
such that $x_j=\+$ for all $j<i$ and $x_j=\-$ otherwise.
We denote the measure concentrated in $(\+\-,i)-$ jump  by $\J_{\+\-}^i.$
 Analogously,
We say that a configuration $x$ is a $(\-\+,i)-${\it jump} if there is position $i$ 
such that $x_j=\-$ for all $j<i$ and $x_j=\+$ otherwise.
We denote the measure concentrated in $(\-\+,i)-$ jump  by $\J_{\-\+}^i.$

\begin{lemma}For each position $j,$ 

(i)If  $\alpha>0$, then
\[
\lim_{t\to\infty}\J_{\-\+}^jF^t=\delta_{\+}. 
\]
(ii)If  $\beta<1$, then
\[
\lim_{t\to\infty}\J_{\+\-}^jF^t=\delta_{\-}. 
\]\label{lem:conJump}
\end{lemma}
{\bf Proof.} First, we will prove item (i). Let $L_1^{\alpha},~L_2^{\alpha},\ldots$ be a sequence of random variable independent identically
distributed, where 
\[
\P(L_1^{\alpha}=1)=\alpha \HB{ and }\P(L_1^{\alpha}=0)=1-\alpha.
\]
Now, we get the simple fact, which can be verified by the Kolmogorov's strong law\cite{Feller}: 

\begin{equation}
  \P\left(\displaystyle\lim_{t\to\infty}\frac{\sum_{N=1}^{t}L^{\alpha}_N}{t}=\alpha\right)=1.
	\label{eq:F1}
\end{equation}

At a informal way, note that by the definition of $F$(see (\ref{eq:transi})),  the random variable 
$\sum_{N=1}^{t}L^{\alpha}_N$ describe the number of new pluses
that has appeared on $\J_{\-\+}^jF^t$. Thus, (\ref{eq:F1})
imply  that the number of pluses goes to infinity  almost surely
and the only way that it can occur is  when $\J_{\-\+}^jF^t$
goes to $\delta_{\+}$ when $t\to\infty.$
{\it Thus, we conclude the proof of item (i).} To prove the item (ii) it is enough
to consider  $L_1^{1-\beta},~L_2^{1-\beta},\ldots$  a sequence of random variable independent identically
distributed, where 
\[
\P(L_1^{1-\beta}=1)=1-\beta \HB{ and }\P(L_1^{1-\beta}=0)=\beta,
\]
and using analog arguments done to prove the item (i) we prove the item (ii).
{\it The lemma \ref{lem:conJump} is proved.}

\begin{lemma}
Lets $x\in\Delta_{\+}$, $y\in\Delta_{\-}$ and $\delta_x$ and $\delta_y$ your respective normalized measures.

(i)If $\beta<1$ then
\[
\lim_{t\to\infty}\delta_xF^t=\delta_{\-}. 
\]

(ii)If $\beta=1$ and $\alpha>0$ then there is position $i$ such that
\[
\lim_{t\to\infty}\delta_xF^t=\J_{\+\-}^i. 
\]

(iii) If $\alpha>0$ then 
\[
\lim_{t\to\infty}\delta_yF^t=\delta_{\+}
\]

(iv)If $\alpha=0$ and $\beta<1$ then there is position $i$ such that
\[
\lim_{t\to\infty}\delta_yF^t=\J_{\-\+}^i. 
\]

(v)If $\beta=1$ and $\alpha=0$ then $\delta_xF=\delta_x$ and $\delta_yF=\delta_y$ 
\label{lem:conisland}
\end{lemma}
{\bf Proof .} The items (ii), (iv)
and (v) are simply. So, we will prove just the items (i)
and  (ii).
Note that given $x$ and $y$ there is value $j$ such that
\[
\delta_x\prec \J_{\+\-}^{j} \HB{ and }\J_{\-\+}^{j}\prec\delta_y, 
\]
By the lemmas \ref{lemm:monoF} and \ref{lem:conJump}
\[
\lim_{t\to\infty}\delta_xF^t\prec\lim_{t\to\infty} \J_{\+\-}^{j}F^t=\delta_{\-} \HB{ for }\beta<1 
\]
and
\[
\delta_{\+}=\lim_{t\to\infty} \J_{\-\+}^{j}F^t\prec\lim_{t\to\infty}\delta_yF^t \HB{ for }\alpha>0. 
\]
As for all $\mu\in{\cal M},~\delta_{\-}\prec\mu\prec\delta_{\+}$, {\it we conclude the proof of lemma 
\ref{lem:conisland}.}

\begin{lemma}
The $\Delta$ is countable.
\label{lem:countable}
\end{lemma}
{\bf Proof.} By the $\Delta$'s definition it is enough to prove that
$\Delta_{\+}$ is countable. It is what we will to do.
Let us define
\[
I_n=\{x\in\Delta_{\+}:~\length(x)=n\},
\]
So,
\[
\Delta_{\+}=\displaystyle \bigcup_{n=1}^{\infty} I_n.
\]
Of course that $I_n$ is countable for all natural value $n,$
then  $\Delta_{\+}$ is countable too. Hence $\Delta$ is countable.
{\it We conclude the proof of the lemma \ref{lem:countable} }.

{\bf Commemt: }The lemma \ref{lem:countable} imply that any $\mu\in\A$
is a finite or  a countably infinite convex combination of $\delta-$measures
of elements of $\Delta$. So from now on, always that we get $\mu\in\A$
we can write 
\[
\mu=\sum_{x\in\Delta}k_x\delta_x,
\]
where $\sum_{x\in\Delta}k_x=1$ and for all $x\in\Delta$ we get $k_x$ are non-negatives 

{\bf Proof of Theorem \ref{TEO10}.} 
We shall prove just the case where $\mu$ is a finite
convex combination of $\delta-$measures. 
The case when $\mu$ is a countably infinite
convex combination of $\delta-$measures is analog. Let $x^1,\ldots,x^N$ islands(of pluses or of minuses) and
$\delta_{x^1},\ldots,\delta_{x^N}$ its respective normalized 
measures. Also, we define $\Omega_{\+}$ the set of island of pluses
and $\Omega_{\-}$ the set of island of minuses. So,
\[
\mu=\sum_{x\in\{x^1,\ldots,x^N\}} k_x\delta_{x}=
\sum_{x\in\{x^1,\ldots,x^N\}\cap\Omega_{\+}} k_x\delta_{x}
+\sum_{ x\in\{x^1,\ldots,x^N\}\cap\Omega_{\-}} k_x\delta_{x},
\]
where $\displaystyle\sum_{x\in\{x^1,\ldots,x^N\}} k_x=1$ and $k_x\ge 0$ for
$x\in\{x^1,\ldots,x^N\}.$ Using the linearity of F(see (\ref{CA})) we get
\[
\mu F^t=\sum_{x\in\{x^1,\ldots,x^N\}\cap\Omega_{\+}} k_x(\delta_{x}F^t)
+\sum_{ x\in\{x^1,\ldots,x^N\}\cap\Omega_{\-}} k_x(\delta_{x}F^t).
\]
Using first the items (i) and (iii) from the lemma \ref{lem:conisland}
and after that 
\[
\sum_{x\in\{x^1,\ldots,x^N\}\cap\Omega_{\+}} k_x=1-\sum_{x\in\{x^1,\ldots,x^N\}\cap\Omega_{\-}} k_x.
\]
 We get, $\mu F^t$ converge to $\pi_{\lambda}$ when $t$ goes to infinity, where $\lambda=\sum_{x\in\{x^1,\ldots,x^N\}\cap\Omega_{\-}} k_x.$ To the particular cases, it is enough to observe that 
if $\mu$ is archipelago of pluses, then
\[
\lambda=\sum_{x\in\{x^1,\ldots,x^N\}\cap\Omega_{\-}} k_x=0.
\]
And if $\mu$ is archipelago of minuses, then
\[
\lambda=\sum_{x\in\{x^1,\ldots,x^N\}\cap\Omega_{\-}} k_x=1.
\]
{\it we conclude the proof of the theorem \ref{TEO10}. }

\begin{figure}
\begin{center}
\small%
\psset{xunit=1cm,yunit=1cm}%
\newrgbcolor{gray1}{.9 .9 .9}
\begin{pspicture}(0,-0.75)(6,6)

\psframe[linestyle = none, fillstyle = solid, fillcolor = gray1](-2,0)(2,4)
\psline[linewidth = 1.75pt, arrowLW=1pt, arrowscale = 2]{o-o}(-2,4)(-2,0)(2,0)(2,4)
\psline[linewidth = 1.75pt, arrowLW=1pt, arrowscale = 2]{o-*}(-2,5.5)(2,5.5)
\psline[linewidth = 1.75pt, arrowLW=1pt, arrowscale = 2]{*-*}(-2,4.7)(-2,4.7)

\rput[b](-2,-0.4){$0$}
\rput[b](2,-0.4){$1$}
\rput[b](0,-0.4){$\alpha$}
\rput[b](2.2,3.6){$1$}
\rput[r](2.4,2){$\beta$}
\rput[r](2.4,5.5){$1$}
\rput(0,2){$\delta_xF^t \rightarrow \delta_{\ominus}$}
\rput(0,5.8){$\delta_xF^t \rightarrow \J_{\oplus\ominus}^i$}
\rput[l](-3.75,4.7){$\delta_xF\rightarrow \delta_x$}
\pspolygon[linestyle = none, fillstyle = solid, fillcolor = gray1](4,4)(4,0)(8,0)
\psline[linewidth = 1.75pt, arrowLW=1pt, arrowscale = 2]{o-o}(4,4)(4,0)(8,0)(8,4)

\rput[b](4,-0.4){$0$}
\rput[b](8,-0.4){$1$}
\rput[b](8.2,3.6){$1$}
\rput[b](6,-0.4){$\alpha$}
\rput[r](8.4,2){$\beta$}

\rput(5.75,1){$\Exp(\tau_x) < \infty $}
\rput(6.25,3){$\Exp(\tau_x) = \infty $}

\end{pspicture}
\end{center}
\caption{{\it The figure on the left side illustrate the results described on the lemma \ref{lem:conisland} about the limit of our process when we started at a measure concentrated at a island of pluses(items (i), (ii) and (v)). Similar illustration is obtained for the items (iii), (iv) and (v) when we started our process at a measure concentrated at a island of minuses. The figure on the right side illustrate the results described on the lemma 
\ref{lem:phaseX} i.e. the behavior of $\Exp (\tau_x)$ when 
$\alpha< 1-\beta$ and  $\alpha\ge 1-\beta$. The illustration of the lemma \ref{lem:phaseY} is similar.}}
\label{Fig:IlusTheo}
\end{figure}
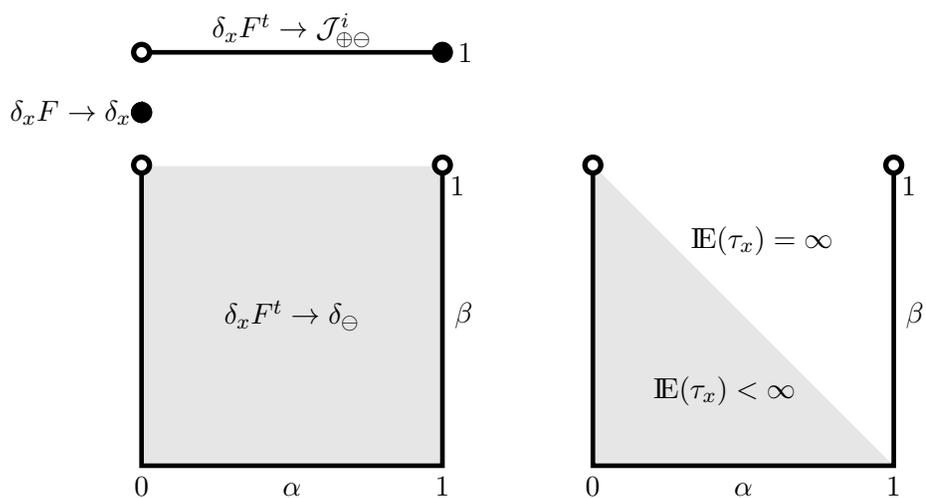
\section{The processes $X$ and $Y$}

Let $X=\{X_t\}_{t=0}^{\infty}$ assuming values  in $\{0,1,2,3,\ldots\}$ where 
\[
\P(X_{t+1}=a+1|X_t=a)=\left\{ 
\begin{array}{ll}
0&\HB{ if }a=0\\
\alpha\beta&\HB{ if }a>0.
\end{array}\right.
\]
and
\[
\P(X_{t+1}=a-1|X_t=a)=\left\{ 
\begin{array}{ll}
0&\HB{ if }a=0\\
(1-\beta)(1-\alpha)&\HB{ if }a>0.
\end{array}\right.
\]
So, 
\[\P(X_{t+1}=a|X_t=a)=1-\P(X_{t+1}=a-1|X_t=a)-\P(X_{t+1}=a+1|X_t=a).\]

We shall denote  
\[
\lim_{t\to\infty}\P(X_t\ge a| X_0=n) \HB{ for every real value }a
\]
 by
\[
\P(X_t\to\infty|X_0=n).
\]

Note that :

\begin{itemize}
	\item If $\alpha=0$ and $0\le\beta<1$ then for all $\epsilon>0,~\P(X_t>\epsilon)\to 0$ when $t\to\infty;$ 
	
	\item If $\beta=0$ and $0<\alpha<1$ then for all $\epsilon>0,~\P(X_t>\epsilon)\to 0$ when $t\to\infty;$
	
	\item If $\alpha=1$ and $0<\beta\le 1$ then $\P(X_t\to\infty|X_0>0)> 0;$  

  \item If $\beta=1$ and $0<\alpha< 1$ then $\P(X_t\to\infty|X_0>0)> 0;$
	
	\item If $\alpha=0$ and $\beta=1$ or $\alpha=1$ and $\beta=0$ then $X_t=X_0$ for all $t>0.$
\end{itemize}

Therefore, we shall not consider those cases  during the proofs of the lemma 
\ref{lem:absorption},\ref{lem:expectation},\ref{lem:absorptionY} and \ref{lem:expectationY}. Thus,
from now on we will take $0<\alpha<1$ and $0<\beta<1.$

 We denote the {\it absorption probability} of our process $X$ hit the state $0$
given that it started on the state $i$ by $h_i.$ We note that
$h_0 = 1$ . The fundamental relationship among the $h_i$'s is the following(see \cite{MC}):
\begin{equation}
\alpha\beta h_{i+1}-((1-\alpha)(1-\beta)+\alpha\beta)h_i+(1-\alpha)(1-\beta)h_{i-1}=0.
\label{eq:absortion}
\end{equation}
For $0<\alpha\le 1$ and $0<\beta\le 1$ we define 
\begin{equation}
\gamma=\frac{(1-\alpha)(1-\beta)}{\alpha\beta}.
\label{eq:gamma}
\end{equation}
\begin{lemma}
(i)If $\alpha\le 1-\beta$ then $h_i=1$ for all $i$;

(ii)If $\alpha >1-\beta$ then $h_i=\gamma^{i}$ for all $i$;

(iii)If $\alpha\le 1-\beta$ then
\[
\P(X_t \to \infty|X_0=i)=0 .
\] 
(iv)If $\alpha>1-\beta$ then
\[
\P(X_t \to \infty|X_0=i)=1-h_i. 
\] 
\label{lem:absorption}
\end{lemma}

{\bf Proof. } Considering $0<\alpha<1$ and $0<\beta<1$,  we get that the general solution 
of (\ref{eq:absortion})
\[
h_i=\left\{
\begin{array}{lc}
A+B\gamma^{i}&\HB{ if }\alpha\not =1-\beta,\\
A+iB&\HB{ if }\alpha =1-\beta.
\end{array}\right.
\] 
where $A$ and $B$ are constants. Using the facts that $h_0=1,~0\le h_i\le 1$
and the general solution of (\ref{eq:absortion})
we can conclude the proof of items (i) and (ii).
The item (iii) is a right consequence of the item (i). Now we shall prove (iv).
Consider a process $X^N=\{{X_t}^N\}_{t=0}^{\infty}$ in $\{0,1,\ldots,N\}$ where 
\[
\P({X_{t+1}}^N =a|{X_t}^N=a)=\P(X_{t+1}=a |X_t=a) \HB{ for all }a\in\{0,1,..,N-1\}
\]
and 
\[
\P({X_{t+1}^N} =N|{X_t}^N=N)=1.
\]
Hence ${X}^N$ has two absorbing states, namely $\{0,N\}.$
When we change the scale we  have the same qualitative behavior
of the process. Thus at  ${X}^N$  for all $a\in\{0,1,..,N-1\}$ we take
\[
\P({X_{t+1}^N} =a+1|{X_t^N}=a)=\tilde{\alpha} \HB{ and }\P({X_{t+1}^N} =a-1|{X_t^N}=a)=\tilde{\beta} 
\]
where
\[
\tilde{\alpha}=\frac{\alpha\beta}{\alpha\beta+(1-\alpha)(1-\beta)} \HB{ and }\tilde{\beta}=\frac{(1-\alpha)(1-\beta)}{\alpha\beta+(1-\alpha)(1-\beta)}
\]
So, $\tilde{\alpha}+\tilde{\beta}=1$ and after rescaling ${X}^N$ became the well-known Gamblers' ruin problem and is well-known
that 
\begin{equation}
\P({X_t}^N=N|X_0^N=i)=
\left\{
\begin{array}{ll}
\frac{i}{N} &\HB{ if }\tilde{\alpha}=\tilde{\beta};\\
\frac{1-(\tilde{\beta}/\tilde{\alpha})^i}{1-(\tilde{\beta}/\tilde{\alpha})^N} &\HB{ if }\tilde{\alpha}\not =\tilde{\beta}.
\end{array}
\right.
\label{eq:gambler}
\end{equation}
Thus, the probability of the gambler became infinitely rich, $\P({X_t}^N=N|X_0^N=i)$ when $N$ goes to $\infty$,
is zero if $\tilde{\beta}\ge \tilde{\alpha}$ and is $1-(\tilde{\beta}/\tilde{\alpha})^i$ if $\tilde{\beta}<\tilde{\alpha}$. But,
\[
\left(\frac{\tilde{\beta}}{\tilde{\alpha}}\right)^i=\gamma^i=h_i \HB{ and }\tilde{\beta}<\tilde{\alpha}\Rightarrow 1-\beta<\alpha.
\]
{\it we  conclude the proof of lemma \ref{lem:absorption}}.

Now, we define another process $Y=\{Y_t\}_{t=0}^{\infty}$ where
\[ \P(Y_{t+1}=a+1|Y_t=a)=\P(X_{t+1}=a-1|X_t=a)\] 
 and 
\[\P(Y_{t+1}=a-1|Y_t=a)=\P(X_{t+1}=a+1|X_t=a).\]
Informally speaking, when $X$ increase $Y$ decrease and when $X$ decrease $Y$ increase.

Let us define the {\it hitting time} of state zero given that we 
stated at the state $i$ by
\[
H_i^X=\inf\{t\ge0: ~X_t=0 \HB{ and } X_0=i \}
\]
 and 
\[
H_i^Y=\inf\{t\ge0: ~Y_t=0 \HB{ and } Y_0=i \}
\]
where the infimum of the empty set is $\infty.$ As common we
denote $\Exp$ the expectation.
So, $\Exp(H_i^X)$ and $\Exp(H_i^Y)$ are  the expected amount 
of time before the process $X$ and $Y$ respectively hits zero,
conditioned  that the process start at $i$.

\begin{lemma}Lets $\beta<1$, $\tilde{\alpha}=\alpha\beta,~\tilde{\beta}=(1-\beta)(1-\alpha)$ and $\gamma$ as defined at (\ref{eq:gamma}).

(i)If $\alpha< 1-\beta$ then
     \[
		\Exp(H_i^X)=
		\left\{
		\begin{array}{ll}
		\displaystyle\frac{1}{\tilde{\beta}}+\frac{\gamma^{-2}}{\tilde{\alpha}(1-\gamma^{-1})}
		&\HB{ if }i=1;\\[15pt]
		\Exp(H_1^X)+\displaystyle\frac{\gamma^{-1}(i-1)}{\tilde{\alpha}(1-\gamma^{-1})}
		&\HB{ if }i> 1.
		\end{array}
		\right.
		\]

(ii)If $\alpha\ge 1-\beta$ then $\Exp(H_i^X)=\infty$ for all $i\ge 1$.
\label{lem:expectation}
\end{lemma}
{\bf Proof. }As a direct consequence of  lemma \ref{lem:absorption} items (ii) and (iv)
we obtain  the item (ii).
The proof of item (i)  is a particular case from the general 
result obtained to a birth and death process(see \cite{Bio} pp:75-77).
{\it Lemma \ref{lem:expectation} is proved.}

{\bf Comment:} The lemma \ref{lem:absorption} show us if $\alpha=1-\beta$ we get $\P(X_t=0|X_0\ge 0)\to 1$ 
when $t\to\infty$, however lemma \ref{lem:expectation} show us that in this same case $\Exp(H_i^X)=\infty$
for all $i>0.$ It can be explained by the fact that the convergence of $X_t$ to zero  when $\alpha=1-\beta$  
occur slowly than when $\alpha<1-\beta.$

 Note that in analog way as proved the lemmas \ref{lem:absorption} and 
\ref{lem:expectation}, we can prove the  lemmas \ref{lem:absorptionY} and
\ref{lem:expectationY}.

\begin{lemma} Let the {\it absorption probability} of our process $Y$ hit the state $0$
given that it started on the state $i$ by $\hat{h}_i.$  

(i)If $\alpha\ge 1-\beta$ then $\hat{h}_i=1$ for all $i$;

(ii)If $\alpha <1-\beta$ then $\hat{h}_i=\gamma^{-i}$ for all $i$;

(iii)If $\alpha\ge 1-\beta$ then
\[
\P(Y_t \to \infty|Y_0=i)=0 .
\] 
(iv)If $\alpha<1-\beta$ then
\[
\P(Y_t \to \infty|Y_0=i)=1-\hat{h}_i . 
\] 
\label{lem:absorptionY}
\end{lemma}

\begin{lemma}Lets $\alpha>0$,  $\tilde{\alpha}=\alpha\beta,~\tilde{\beta}=(1-\beta)(1-\alpha)$ and
$\gamma$ as defined at (\ref{eq:gamma})

(i)If $\alpha\le 1-\beta$ then $\Exp(H_i^Y)=\infty$ for all $i\ge 1$.

(ii)If $\alpha> 1-\beta$ then
     \[
		\Exp(H_i^Y)=
		\left\{
		\begin{array}{ll}
		\displaystyle\frac{1}{\tilde{\alpha}}+\frac{\gamma^2}{\tilde{\beta}(1-\gamma)}
		&\HB{ if }i=1;\\[15pt]
		\Exp(H_1^Y)+\displaystyle\frac{\gamma(i-1)}{\tilde{\beta}(1-\gamma)}
		&\HB{ if }i> 1.
		\end{array}
		\right.
		\]

\label{lem:expectationY}
\end{lemma}

\section{Proof of theorems \ref{TEO20}, \ref{TEO30} and \ref{TEO40}}

Given $x\in\Delta_{\+}$, we denote the minimum value $i$ such that $x_i=\+$
by $i_{min}$ and the maximum value $i$ such that $x_i=\+$
by $i_{max}.$  Thus, we shall define the following configurations
\[
(\underline{x})_i=\left\{
\begin{array}{ll} 
\+&\HB{ if }i=i_{max} ; \\ 
\-&\HB{ otherwise.}
\end{array}
\right.
\HB{ and }
 (\overline{x})_i=\left\{
\begin{array}{ll} 
\+&\HB{ if }i_{min} \le i \le i_{max}  ; \\ 
\-&\HB{ otherwise.}
\end{array}
\right. 
\]
Note that $\overline{x},~\underline{x}\in\Delta_{\+}$  and
\[
\underline{x}\prec x\prec \overline{x}. 
\]

We will consider island of pluses where $x=\overline{x}$. Thus, 
 there are positions $i<j$ such that $x_k=\+$ if $i<k<j$ 
and $x_k=\-$ otherwise and for those same positions $i$ and $j$
we get $\length(x)=j-i-1.$ If $n=1$ then
$\underline{x}=x=\overline{x}.$  
Take $x=\overline{x}$, we will associate our process acting in $\delta_x$  with $X$.

\begin{figure}[htp]
\begin{center}
\setlength{\unitlength}{0.8cm}
\begin{picture}(5,2)
\put(1,0){\line(1,0){5}}
\put(1,0){\circle*{0.20}}
\put(2,0){\circle*{0.20}}
\put(3,0){\circle*{0.20}}
\put(4,0){\circle*{0.20}}
\put(5,0){\circle*{0.20}}
\put(6,0){\circle*{0.20}}
\put(0,0){\ldots}
\put(6.3,0){\ldots}

\put(1,-0.5){$x_1$}
\put(2,-0.5){$x_2$}
\put(3,-0.5){$x_3$}
\put(4,-0.5){$x_4$}
\put(5,-0.5){$x_5$}
\put(6,-0.5){$x_6$}

\put(0.8,0.3){$\ominus$}
\put(1.8,0.3){$\ominus$}
\put(2.8,0.3){$\ominus$}
\put(3.8,0.3){$\+$}
\put(4.8,0.3){$\+$}
\put(5.8,0.3){$\ominus$}
\put(0,0.3){\ldots}
\put(6.3,0.3){\ldots}

\put(0.8,0.7){$\ominus$}
\put(1.8,0.7){$\ominus$}
\put(2.8,0.7){$\+$}
\put(3.8,0.7){$\+$}
\put(4.8,0.7){$\ominus$}
\put(5.8,0.7){$\ominus$}
\put(0,0.7){\ldots}
\put(6.3,0.7){\ldots}

\put(0.8,1.1){$\ominus$}
\put(1.8,1.1){$\+$}
\put(2.8,1.1){$\+$}
\put(3.8,1.1){$\+$}
\put(4.8,1.1){$\ominus$}
\put(5.8,1.1){$\ominus$}
\put(0,1.1){\ldots}
\put(6.3,1.1){\ldots}

\put(0.8,1.5){$\ominus$}
\put(1.8,1.5){$\+$}
\put(2.8,1.5){$\+$}
\put(3.8,1.5){$\ominus$}
\put(4.8,1.5){$\ominus$}
\put(5.8,1.5){$\ominus$}
\put(0,1.5){\ldots}
\put(6.3,1.5){\ldots}

\put(0.8,1.9){$\ominus$}
\put(1.8,1.9){$\+$}
\put(2.8,1.9){$\+$}
\put(3.8,1.9){$\ominus$}
\put(4.8,1.9){$\ominus$}
\put(5.8,1.9){$\ominus$}
\put(0,1.9){\ldots}
\put(6.3,1.9){\ldots}

\put(7.3,0){$X_0=2$}
\put(7.3,0.5){$X_1=2$}
\put(7.3,1){$X_2=3$}
\put(7.3,1.5){$X_3=2$}
\put(7.3,2){$X_4=2$}
\end{picture}
\vspace{-0.3cm}
\end{center}
\caption{{\it Here we illustrate a fragment of our process, which occur with positive probability.
The initial configuration is a island of pluses, $x$, where $x=\overline{x}$, the length of the
island is $2$ and $i_{min}=4$ and $i_{max}=5$. Also, $i_0=3 $ and $j_0=6$; $i_1=2 $ and $j_1=5$;
$i_2=1 $ and $j_2=5$; $i_3=1 $ and $j_3=4$ and $i_4=1 $ and $j_4=4$. Also, on the right side we show
the correspondent values assumed by the $X$ process. }}
\label{coupling}
\end{figure}
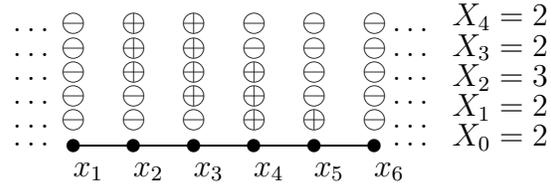

Give a island of pluses $x$ where $x=\overline{x}$ and respective 
normalized measure concentrated in $x$, 
$\delta_x.$ There are positions $i_0<j_0$  such that
$x_{i_0}=x_{j_0}=\-$ and $x_k=\+$ if $i_0<k<j_0$. We assume
$X_0=j_0-i_0-1$, note that $X_0=\length(x)$,
 what is the number of consecutive pluses between the positions $i_0$
and $j_0$. Also we define $X_t=j_t-i_t-1$, where  
 the random variables $i_t$ and $j_t$, ($i_t<j_t$),  are defined as follows(see Figure \ref{coupling}):
\[
\P(i_t=i_{t-1}-1, ~j_t=j_{t-1})=
\left\{
\begin{array}{ll}
0& \HB{ if }j_{t-1}=i_{t-1}+1;\\
\theta(\+|\-\+)\theta(\+|\+\-)& \HB{ otherwise. }
\end{array}
\right.
\]
\[
\P(i_t=i_{t-1}-1, ~j_t=j_{t-1}-1)=
\left\{
\begin{array}{ll}
0& \HB{ if }j_{t-1}=i_{t-1}+1;\\
\theta(\+|\-\+)\theta(\-|\+\-)& \HB{ otherwise. }
\end{array}
\right.
\]
\[
\P(i_t=i_{t-1}, ~j_t=j_{t-1})=
\left\{
\begin{array}{ll}
1& \HB{ if }j_{t-1}=i_{t-1}+1;\\
\theta(\-|\-\+)\theta(\+|\+\-)& \HB{ otherwise. }
\end{array}
\right.
\]
\[
\P(i_t=i_{t-1}, ~j_t=j_{t-1}-1)=
\left\{
\begin{array}{ll}
0& \HB{ if }j_{t-1}=i_{t-1}+1;\\
\theta(\-|\-\+)\theta(\-|\+\-)& \HB{ otherwise. }
\end{array}
\right.
\]
$\theta(.|.)$ is  the probability transitions of our process (\ref{eq:transi}). Note that 
$i_t$ and $j_t$ describe the probability of the length of the island of pluses: increase,  decrease or stay.
If $x$ is a island of pluses $\delta_xF^t$ will be a measure 
concentrated at a island of pluses for each natural value $t$. 
 
Now, it is easy to conclude:
\[
\begin{array}{lcl}
\P(X_t=a+1|X_{t-1}=a)&=&\P(i_t=i_{t-1}-1, ~j_t=j_{t-1});\\
\P(X_t=a-1|X_{t-1}=a)&=&\P(i_t=i_{t-1}, ~j_t=j_{t-1}-1);\\
\P(X_t=a|X_{t-1}=a)&=&\P(i_t=i_{t-1}, ~j_t=j_{t-1})+\P(i_t=i_{t-1}-1, ~j_t=j_{t-1}-1).
\end{array}
\]
Where $a=j_{t-1}-i_{t-1}-1$. Thus, we have conclude the task to 
associate our process acting in $\overline{x}$ with $X$.

Lets $x\in\Delta_{\+}$  and $y\in\Delta_{\-}$ , we define the random variables
\[
\tau_x=\inf\{t\ge 0:\delta_x F^t=\delta_{\-}\} \HB{ and }\tau_y=\inf\{t\ge 0:\delta_y F^t=\delta_{\+}\}
\]
the infimum of the empty set is $\infty.$

\begin{lemma}Lets $\beta<1$, $x\in\Delta_{\+}$  and  $\delta_x$ your respective normalized measures . 

(i) If $\alpha< 1-\beta$ then $\Exp(\tau_x)<\infty;$

(ii) If $\alpha\ge 1-\beta$ then $\Exp(\tau_x)=\infty;$

\label{lem:phaseX}
\end{lemma}

{\bf Proof .} For any island of pluses, $x$, we get
\[
\delta_{\underline{x}}\prec\delta_{x}\prec\delta_{\overline{x}}.
\]
So, by the lemma \ref{lemm:monoF} for any natural value $t$ 
\begin{equation}
\delta_{\underline{x}}F^t\prec\delta_{x}F^t\prec\delta_{\overline{x}}F^t.
\label{eq:ineq}
\end{equation}
Hence, 
\begin{equation}
\Exp(\tau_{\underline{x}})\le\Exp(\tau_{x})\le\Exp(\tau_{\overline{x}}).
\label{eq:expineq}
\end{equation}
 We showed previously in this section
the association between the process $X$ and the evolution of the length of the island.
That association  we shall use in this proof.
Now, we shall prove the item (ii).
Note that
\[
\Exp(\tau_{\underline{x}})=\Exp(H_1^X).
\]
By the lemma \ref{lem:expectation} item (ii) we get 
if $\alpha\ge 1-\beta$ then $\Exp(H_1^X)=\infty$, then $\Exp(\tau_{\underline{x}})=\infty$.
Thus, using (\ref{eq:expineq}) we get  $\Exp(\tau_{{x}})=\infty$.{\it we conclude the proof 
of (ii)}.
Now, we shall prove the item (i). 
Let $\overline{x}$ be a island of plus whose \length($\overline{x}$)$=n$, so
\[
\Exp(\tau_{\overline{x}})=\Exp(H_n^X).
\]
By the lemma \ref{lem:expectation} item (i) we get 
if $\alpha< 1-\beta$ then $\Exp(H_n^X)$ is finite, then $\Exp(\tau_{\underline{x}})$ is finite.
Thus, using (\ref{eq:expineq}) we conclude that  $\Exp(\tau_{{x}})$ is finite.
{\it The  lemma \ref{lem:phaseX} is proved}.

\begin{lemma}Lets $\alpha>0$, $y\in\Delta_{\-}$ and  $\delta_y$ your respective normalized measures. 

(i) If $\alpha> 1-\beta$ then $\Exp(\tau_y)<\infty;$

(ii) If $\alpha\le 1-\beta$ then $\Exp(\tau_y)=\infty;$

\label{lem:phaseY}
\end{lemma}

{\bf Proof.} The proof is analog to the proof of lemma \ref{lem:phaseX}.
It is enough  associate our process with $Y$, what can be done naturally by take a island of 
minuses. The $Y$ associated in this way will describe the probability of  the length of the island:
decrease, increase and stay the same.  

Also, we need to define: given a island of minus, $y$, 
we  denote the minimum value $i$ such that $y_i=\-$
by $i_{min}$ and the maximum value $i$ such that $y_i=\-$
by $i_{max}.$  Thus, we shall denote the following configurations
\[
(\underline{y})_i=\left\{
\begin{array}{ll} 
\-&\HB{ if }i=i_{max} ; \\ 
\+&\HB{ otherwise.}
\end{array}
\right.
\HB{ and }
 (\overline{y})_i=\left\{
\begin{array}{ll} 
\-&\HB{ if }i_{min} \le i \le i_{max}  ; \\ 
\+&\HB{ otherwise.}
\end{array}
\right. 
\]
Therefore,
\[
\overline{y}\prec y\prec \underline{y}. 
\]
Thus,
\[
\Exp(\tau_{\underline{y}})\le\Exp(\tau_{y})\le\Exp(\tau_{\overline{y}}).
\]
So, using  the lemma \ref{lem:expectationY} {\it we conclude the  lemma \ref{lem:phaseY}.}

 On the cases when $\Exp(\tau_x)$ and $\Exp(\tau_y)$ are finite, through the lemmas \ref{lem:expectation}
and \ref{lem:expectationY} we can  obtain  a estimation of the mean time for the island `` disappear''. 
In this direction,  we will prove the lemma \ref{lem:ineexptime}.
\begin{lemma}Let $\gamma$
as defined at (\ref{eq:gamma}).
Given $\alpha$ and $\beta$ there are constants $k_1$ and $k_2$ such that:

(i)Let $\beta<1$ be. If $\alpha<1-\beta$ then 
\[
k_1+k_2 \gamma\le\Exp(\tau_x)\le k_1+(\length(x)-1) k_2 \HB{ for all }x\in\Delta_{\+}. 
\]
(ii)Let $\alpha>0$ be.If $\alpha>1-\beta$ then 
\[
k_2+k_1 \gamma\le\Exp(\tau_y)\le k_2+(\length(y)-1) k_1 \HB{ for all  }y\in\Delta_{\-}. 
\]
\label{lem:ineexptime}
\end{lemma}  

{\bf Proof.} We shall prove (i). 
By the lemma \ref{lem:expectation}
\[
\Exp(H^X_n)=
\left\{
\begin{array}{ll}
\frac{1}{(1-\beta)(1-\alpha)}+\frac{\gamma^{-2}}{\alpha\beta(1-\gamma^{-1})}&\HB{ for }n=1;\\
\frac{1}{(1-\beta)(1-\alpha)}+\frac{(n-1)\gamma^{-1}}{\alpha\beta(1-\gamma^{-1})}&\HB{for }n>1.
\end{array}
\right.
\]
So, 
\begin{equation}
 k_1+\gamma k_2 =\Exp(H^X_1)\HB{ and }\Exp(H^X_n)= k_1+(n-1) k_2
\label{eq:ineqestimation}
\end{equation}
where $k_1=((1-\beta)(1-\alpha))^{-1}$ and $k_2=\gamma^{-1}/(\alpha\beta(1-\gamma^{-1})).$ 

Now let us consider $x$ a island of pluses, of course that  $x$ and $\overline{x}$ are islands with the same  length.  Using (\ref{eq:expineq}) we get
\[
\Exp(\tau_{\underline{x}})\le\Exp(\tau_{x})\le\Exp(\tau_{\overline{x}}).
\] 
Also, we know that $\Exp(\tau_{\underline{x}})=\Exp(H_1^X)$ and for \length($\overline{x}$)$=n$ we get$\Exp(\tau_{\overline{x}})=\Exp(H_n^X)$.
Thus using (\ref{eq:ineqestimation}) {\it  we conclude the proof of item (i)}. The proof of $(ii)$ is analog.{\it The lemma \ref{lem:ineexptime} is proved} .

We say that our operator $F$ is {\it eroder of island of  pluses in mean linear time}(respectively {\it eroder of island of minuses in mean linear time}) if  fixed $\alpha$ and $\beta$ there is constant
$k$ such that
\[
\Exp(\tau_x)\le k(1+\length(x)) \HB{ for all } x\in\Delta_{\+}.
\]
(Respectively for all $x\in\Delta_{\-}$).
Here we are using the name eroder different of that used at \cite{Stable,Galperin,Petri}.
There the name eroder  was used for deterministic operators.

\begin{lemma} Lets $\alpha>0$ and $\beta<1$ be.

(i)If $\alpha<1-\beta$ then $F$ is eroder of island of pluses in mean linear time; 

(ii)If $\alpha>1-\beta$ then $F$ is  eroder of island of minuses in mean linear time;
\label{lem:reoderlinear}
\end{lemma}  
 
{\bf Proof .} Straight from the lemma \ref{lem:ineexptime}.

On the theorems \ref{TEO20},\ref{TEO30} and \ref{TEO40}, 
we shall prove just the case where $\mu$ is a finite
convex combination of $\delta-$measures. 
The case when $\mu$ is a countably infinite
convex combination of $\delta-$measures is analog.

{\bf Proof of the theorem \ref{TEO20}.}  

Let $\mu$ a archipelago of pluses. So,
\[
\mu=\sum_{x\in\{x^1,\ldots,x^N\}}k_x\delta_x,
\]
where $\sum_{i=1}^Nk_{x^i}=1;~k_{x^1},\ldots,k_{x^N}$ are positives  and $x^1,\ldots,x^N$ are island of pluses. 
Note that by the theorem \ref{TEO10} and $\mu$ definition 
\[
\begin{array}{lcl}
\tau_{\mu}&=&\inf\{t \ge 0:~\mu F^t=\delta_{\-}\}\\
&=&\inf\{t \ge 0:~k_{x^1} (\delta_{x^1}F^t)+\ldots+k_{x^N} (\delta_{x^N}F^t)=\delta_{\-}\}\\
&=&\inf\{t \ge 0:~ (\delta_{x^1}F^t)=\ldots= (\delta_{x^N}F^t)=\delta_{\-}\}.
\end{array}
\]
By the lemma \ref{lem:phaseX} if $\alpha\ge1-\beta$ we get
\[
\Exp(\tau_{x^i})=\infty\HB{ for all }i=1,\ldots,N, 
\]
what imply $\Exp(\tau_{\mu})=\infty.$
Also, using
the lemma \ref{lem:phaseX} if $\alpha<1-\beta$ we get
\[
\Exp(\tau_{x^i})<\infty\HB{ for all }i=1,\ldots,N, 
\]
what imply $\Exp(\tau_{\mu})<\infty.$
{\it Thus we have conclude the proof of the theorem \ref{TEO20}}.

{\bf Proof of theorem \ref{TEO30}.} The proof is analog to the proof of the 
theorem \ref{TEO20}. We just need to use the  lemma \ref{lem:phaseY} and   
  that  
$\tau_{\mu}=\inf\{t\ge 0:~\mu F^t=\delta_{\oplus}\}.$

{\bf Proof of theorem \ref{TEO40}.} First we shall prove (A.\ref{TEO40}). 
If $\mu$ is a archipelago of pluses then
\[
\mu=\sum_{x\in\{x^1,\ldots,x^N\}}k_x\delta_x,
\]
where $\sum_{i=1}^Nk_{x^i}=1;~k_{x^1},\ldots,k_{x^N}$ are positives  and $x^1,\ldots,x^N$ are island of pluses. 
By the theorem \ref{TEO10} and $\mu$ definition 
\[
\tau_{\mu}=\inf\{t \ge 0:~\mu F^t=\delta_{\-}\}
=\inf\{t \ge 0:~ (\delta_{x^1}F^t)=\ldots= (\delta_{x^N}F^t)=\delta_{\-}\}.
\]
Hence, 
\[
\Exp(\tau_{\mu})\le\max\{\Exp(\tau_{x^i}),~i=1,\ldots,N\}.
\]
By the lemma \ref{lem:reoderlinear} item (i) given $\alpha$ and $\beta$ such that $\alpha<1-\beta$
 there is constant $k$ such that 
\[
\Exp(\tau_{x^i})\le k(1+\length(x^i))\HB{ for all }i=1,\ldots,N.
\]
 therefore for that same constant $k$
\[
\begin{array}{lcl}
\max\{\Exp(\tau_{x^i}),~i=1,\ldots,N\}&\le& k\left(1+\max\{\length(x^i):~i=1,\ldots,N\}\right)\\[4pt]
&=&k(1+\giant(\mu)).
\end{array}
\]
Thus, we have conclude the proof of (A.\ref{TEO40}). 
 The proof of (B.\ref{TEO40}) is analog.{\it we conclude the proof of theorem \ref{TEO40}}

The lemma \ref{exp:archipelagod} describe to us what occur with $\Exp(\tau_{\mu})$
when $\mu\in\A\setminus(\A_{\-}\cup\A_{\+})$ i.e when $\mu$ is not a archipelago
of pluses or minuses.

\begin{lemma}Lets $\alpha>0$ and $\beta<1$ be.
If $\mu\in\A\setminus(\A_{\-}\cup\A_{\+})$ then $\Exp(\tau_{\mu})=\infty.$
\label{exp:archipelagod}
\end{lemma}
{\bf Proof.} As considered on the theorems \ref{TEO20},\ref{TEO30} and \ref{TEO40} we 
will prove just for the case when $\mu$ is a finite convex combination of $\delta-$measures.
If $\mu\in\A\setminus(\A_{\-}\cup\A_{\+})$ then 
\[
\mu=\mu_x+\mu_y,
\]
where
\[
\mu_x=\sum_{x\in\{x^1,\ldots,x^i\}}k_{x}\delta_x,~~\mu_y=\sum_{y\in\{y^{i+1},\ldots,y^N\}}k_{y}\delta_y,
\]
  $x^1,\ldots,x^i$ belongs to $\Delta_{\+}$ and $y^{i+1},\ldots,y^N$ belongs to $\Delta_{\-}.$
Note that
\[
\begin{array}{lll}
\tau_{\mu}&=&\inf\{t\ge 0:\delta_{x^1}F^t=\ldots=\delta_{x^i}F^t=\delta_{\-} \hbox{ and }\delta_{y^{i+1}}F^t=\ldots=\delta_{y^N}F^t=\delta_{\+}\}.\\
&\ge&\inf\{t\ge 0:\delta_{y^{i+1}}F^t=\ldots=\delta_{y^N}F^t=\delta_{\+}\}\\
&=&\tau_{\mu_y}.
\end{array}
\]
Also
\[
\tau_{\mu}\ge\inf\{t\ge 0:\delta_{x^1}F^t=\ldots=\delta_{x^i}F^t=\delta_{\-} \}=\tau_{\mu_x}
\]
So, $\Exp(\tau_{\mu})\ge\Exp(\tau_{\mu_y})$ and $\Exp(\tau_{\mu})\ge\Exp(\tau_{\mu_x})$. 
By the lemmas \ref{lem:phaseX} and \ref{lem:phaseY}: if $\alpha< 1-\beta$ then  $\Exp(\tau_{y^j})=\infty$ for all $j=i+1,\ldots,N$. So $\Exp(\tau_{\mu_y})=\infty$, thus $\Exp( \tau_{\mu})=\infty.$
Another hand,   if $\alpha\ge 1-\beta$ then $\Exp(\tau_{x^j})=\infty$ for all $j=1,\ldots,i$. 
So $\Exp(\tau_{\mu_x})=\infty$, thus $\Exp(\tau_{\mu})=\infty.$
{\it We conclude the proof of lemma \ref{exp:archipelagod}}.

\subsection{Finite space}

Any cellular automaton may have infinite space $\Z$
or finite space $\Z_n$ - {\it the set of remainders modulo} 
$n$, where $n$ is an arbitrary
natural number.
In this  case we have
a finite Markov chain which is ergodic except
degenerate cases. But the speed of convergence
may be very different for different values of parameters.
Note that  our process at finite space is
closer to computer simulation.

\begin{figure}[ht]
\begin{center}
\setlength{\unitlength}{0.8cm}
\begin{picture}(5,1.8)
\put(0,0){\line(1,0){4.2}}
\put(0,1){\line(1,0){4.2}}
\put(1,0){\line(0,1){1}}
\put(0,0){\line(0,1){1}}
\put(2,0){\line(0,1){1}}
\put(3,0){\line(0,1){1}}
\put(4.2,0){\line(0,1){1}}
\qbezier(4.2, 0.5)(6.5, -0.5)(2.1, -0.6)
\qbezier(2.1, -0.6)(-2.3, -0.5)(0, 0.5)
\put(0.2,0.2){$C_0$}
\put(1.2,0.2){$C_1$}
\put(2.2,0.2){\ldots}
\put(3.0,0.2){$C_{n-1}$}
\end{picture}
\end{center}
\caption{A circular $C$ with $|C|=n.$}
\label{circular}
\end{figure}
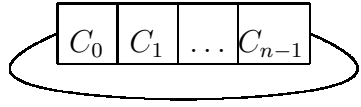

To our finite cellular automata, let us consider the 
set of states $\Omega_n=\{\+,\-\}^{\Z_n}$.
 Elements of $\Omega_n$ we
call {\it circulars}. 
The circulars are finite sequences of pluses $\+$ and minuses $\-$, 
but now we imagine these sequences to have circular form.
We denote by $|C|$ the number of components in a circular $C$
(see figure \ref{circular} where $|C|=n$).
Also we we shall denote the circular whose all the components are equal to $\oplus$,
$C_{\+}$ and the circular whose all the components are equal to $\ominus$,
$C_{\-}$.

Note that $\Omega_n$ has $2^n$ circulars.
We denote by ${\cal M}_{\Omega_n}$ the set of distributions in $\Omega_n.$
 The circular obtained at time 
$t$ was denoted by $C^t$ and its $i$-th components were denoted by 
$C^t_i,$ where $i=0,\ldots,|C^t|-1.$
We consider the circulars $C^t$ as representations of measures $\mu^t\in{\cal M}_{\Omega_n},$ 
so the sequence $C^0,~C^1,~ C^2\ldots$, is a trajectory of 
some random process $\mu^0,\mu^1,\mu^2,\ldots.$

Note that different of infinity space, our process with finite space
if started with a configuration $C$ different of $C_{\+}$ and $C_{\-}$  
we get
\[
\P(\exists~ t_0:~t>t_0 \HB{ imply }C^t = C_{\+})>0 
\]
 and 
\[
\P(\exists~ t_0:~t>t_0 \HB{ imply }C^t = C_{\-})>0.
\]

Given  $C$, whose  $C^0=C$ and 
$|C|=n$, we define
\[
\tau_{C}^n=\inf\{t\ge 1: C^t=C_{\-}\HB{ or } C^t=C_{\+}\}.
\]
Let $X^n$ the process defined to obtain  (\ref{eq:gambler})( Gambler' ruin problem at $\{0,\ldots,n\}$) we define
\[
H^{X^n}_i=\inf\{t\ge0 :~X_t^n\in\{0,n\} \HB{ and }X_0^n=i\}.
 \]  
Using theorem 1.3.5 in \cite{MC} and the probabilities transition of $X^n$  we get
\[
\Exp(H^{X^n}_i)=
\left\{
\begin{array}{ll}
\displaystyle\frac{n(1-\gamma^i)+i(\gamma^n-1)}{(1-\gamma^n)(1-\alpha-\beta)}&\HB{for }\gamma\not =1;\\[10pt]
\displaystyle\frac{in-i^2 }{2(1-\alpha)\alpha}&\HB{for }\gamma = 1.
\end{array}
\right.
\]
where $\gamma=((1-\alpha)(1-\beta))/\alpha\beta$. Hence, for $0< i<n$
\[
\Exp(H_1^{X^n})\le \Exp(H_i^{X^n})\le\left\{
\begin{array}{ll}
n&\HB{for }\gamma\not =1;\\[15pt]
n^2&\HB{for }\gamma = 1.
\end{array}
\right.
\]

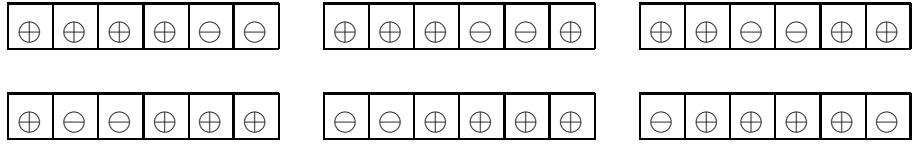
\begin{figure}[ht]
\begin{center}
\setlength{\unitlength}{0.6cm}
\begin{picture}(5,3)
\put(7,2){\line(1,0){6}}
\put(7,3){\line(1,0){6}}
\put(8,2){\line(0,1){1}}
\put(7,2){\line(0,1){1}}
\put(9,2){\line(0,1){1}}
\put(10,2){\line(0,1){1}}
\put(11,2){\line(0,1){1}}
\put(12,2){\line(0,1){1}}
\put(13,2){\line(0,1){1}}
\put(7.2,2.2){$\+$}
\put(8.2,2.2){$\+$}
\put(9.2,2.2){$\ominus$}
\put(10.2,2.2){$\ominus$}
\put(11.2,2.2){$\+$}
\put(12.2,2.2){$\+$}

\put(0,2){\line(1,0){6}}
\put(0,3){\line(1,0){6}}
\put(1,2){\line(0,1){1}}
\put(0,2){\line(0,1){1}}
\put(2,2){\line(0,1){1}}
\put(3,2){\line(0,1){1}}
\put(4,2){\line(0,1){1}}
\put(5,2){\line(0,1){1}}
\put(6,2){\line(0,1){1}}
\put(0.2,2.2){$\+$}
\put(1.2,2.2){$\+$}
\put(2.2,2.2){$\+$}
\put(3.2,2.2){$\ominus$}
\put(4.2,2.2){$\ominus$}
\put(5.2,2.2){$\+$}

\put(-7,2){\line(1,0){6}}
\put(-7,3){\line(1,0){6}}
\put(-6,2){\line(0,1){1}}
\put(-7,2){\line(0,1){1}}
\put(-5,2){\line(0,1){1}}
\put(-4,2){\line(0,1){1}}
\put(-3,2){\line(0,1){1}}
\put(-2,2){\line(0,1){1}}
\put(-1,2){\line(0,1){1}}
\put(-6.8,2.2){$\+$}
\put(-5.8,2.2){$\+$}
\put(-4.8,2.2){$\+$}
\put(-3.8,2.2){$\+$}
\put(-2.8,2.2){$\ominus$}
\put(-1.8,2.2){$\ominus$}

\put(7,0){\line(1,0){6}}
\put(7,1){\line(1,0){6}}
\put(8,0){\line(0,1){1}}
\put(7,0){\line(0,1){1}}
\put(9,0){\line(0,1){1}}
\put(10,0){\line(0,1){1}}
\put(11,0){\line(0,1){1}}
\put(12,0){\line(0,1){1}}
\put(13,0){\line(0,1){1}}
\put(7.2,0.2){$\ominus$}
\put(8.2,0.2){$\+$}
\put(9.2,0.2){$\+$}
\put(10.2,0.2){$\+$}
\put(11.2,0.2){$\+$}
\put(12.2,0.2){$\ominus$}

\put(0,0){\line(1,0){6}}
\put(0,1){\line(1,0){6}}
\put(1,0){\line(0,1){1}}
\put(0,0){\line(0,1){1}}
\put(2,0){\line(0,1){1}}
\put(3,0){\line(0,1){1}}
\put(4,0){\line(0,1){1}}
\put(5,0){\line(0,1){1}}
\put(6,0){\line(0,1){1}}
\put(0.2,0.2){$\ominus$}
\put(1.2,0.2){$\ominus$}
\put(2.2,0.2){$\+$}
\put(3.2,0.2){$\+$}
\put(4.2,0.2){$\+$}
\put(5.2,0.2){$\+$}

\put(-7,0){\line(1,0){6}}
\put(-7,1){\line(1,0){6}}
\put(-6,0){\line(0,1){1}}
\put(-7,0){\line(0,1){1}}
\put(-5,0){\line(0,1){1}}
\put(-4,0){\line(0,1){1}}
\put(-3,0){\line(0,1){1}}
\put(-2,0){\line(0,1){1}}
\put(-1,0){\line(0,1){1}}
\put(-6.8,0.2){$\+$}
\put(-5.8,0.2){$\ominus$}
\put(-4.8,0.2){$\ominus$}
\put(-3.8,0.2){$\+$}
\put(-2.8,0.2){$\+$}
\put(-1.8,0.2){$\+$}
\end{picture}
\end{center}
\caption{{\it All the six circulares are blocks, whose $|B|=6$ and $l(B)=4.$ Actually, fixing one of the blocks we can obtain all of the others using  shift .}}
\label{blocks}
\end{figure}

To prove the lemma \ref{lem:phaseX}, we associate the process $X$ with the evolution of island of pluses.
At similar way we can associate $X^n$ with the evolution of two kinds of circulars, $B\not\in\{C_{\-},C_{+}\}$, defined as follows:  there are $0\le i<j\le n-1$ such that
(i) $B_k=\oplus$ for all $i<k<j$ and $B_k=\ominus$ otherwise or (ii) $B_k=\ominus$ for all $i<k<j$ and $B_k=\ominus$ otherwise. We will call $B$ {\it blocks} and we will denote the number of pluses in $B,~l(B)$ (see Figure \ref{blocks}).  

Note that using the association between  any block $B$ and $X^n$, 
\[ \Exp(H_1^{X^n})\le\Exp(\tau_{B}) =\Exp(H^{X^n}_{l(B)}).\]
 
Of course that any measure $\mu\in{\M}_{\Omega_n}$ is a finite convex combination of $\delta-$measures
concentrated at circulars, $\delta_C$. Thus, if 
\begin{equation}
\mu=\sum_{B\in\Omega_n}k_B\delta_B,
\label{eq:blocksmu}
\end{equation}
where $B$ are blocks, $\sum_{B\in\Omega_n}k_B =1$ and $k_B$ are non-negative real values and
\[
\tau_{\mu}^n=\inf\{t\ge 0:~\mu^t=C_{\-} \HB{ or }\mu^t=C_{\+}\},
\]
 then
\[
\Exp(\tau_{\mu}^n)=\max\{\Exp(\tau_{B}^n):~k_B>0\} \le\max\{\Exp(H^{X^n}_i):~i=0,\ldots,n\}.
\]
Therefore, if $\mu$ is of the form (\ref{eq:blocksmu}),  then 
$\Exp(\tau_{\mu}^n)$ is  of the order\footnote{As well-known $f(n)=O(g(n))$ if only if there is constants $c$ and $n_0$ such that $|f(n)|\le c |g(n)|$ for all $n>n_0.$
} $O(n)$ for $\gamma\not =1$ and $O(n^2)$ for $\gamma =1.$ It means that when we perform a computer simulation of this process, assuming initially
a circular whose $|B|=n$, we will wait at average, no more that $n$ time steps for $\gamma\not =1$ and $n^2$ time steps for $\gamma =1$ to our process achieve  one absorption state.
At another side,
\[
\Exp(\tau_{\mu}^n)\ge\min\{\Exp(H^{X^n}_i):~i=0,\ldots,n\}\ge\Exp(H^{X^n}_1).
\]
But,
\[
\Exp(H^{X^n}_1)=
\left\{\begin{array}{ll}
k_{(\alpha,\beta)}^1n-k_{(\alpha,\beta)}^2 & \HB{ if }\gamma\not =1;\\[10pt]
k_{\alpha}(n-1)&\HB{ if }\gamma =1.
\end{array}
\right.
\]
where $k_{\alpha}=1/(2(1-\alpha)\alpha),~k_{(\alpha,\beta)}^1=(1-\gamma)/((1-\gamma^n)(1-\alpha-\beta))$ and
$k_{(\alpha,\beta)}^2=1/(1-\alpha-\beta)$. Thus, we can conclude
\[
k_{(\alpha,\beta)}^1n-k_{(\alpha,\beta)}^2\le \Exp(\tau_{\mu}^n)\le n\HB{ if }\gamma\not =1
\]
and
\[
k_{\alpha}(n-1)\le\Exp(\tau_{\mu}^n)\le n^2\HB{ if }\gamma =1.
\]
Hence, given values $\alpha$ and $\beta$ if $\gamma\le 1$ i. e. $\alpha\ge 1-\beta$ then $\Exp(\tau_{\mu}^n)$ goes to infinity when $n\to\infty.$ What agree with the results on the theorem \ref{TEO20}. If $\gamma>1$ we can not conclude anything about  the behavior of $\Exp(\tau_{\mu}^n)$ when $n\to\infty$. 

\section*{Acknowledgments}
A. D. Ramos  was supported by CNPQ and CAPES, Processo: BEX 2176/130.

\label{LastPage}

\end{document}